# A Logical Temporal Relational Data Model

Nadeem MAHMOOD[1], Aqil BURNEY[2] and Kamran AHSAN[3]

[1,2] Department of Computer Science (UBIT), University of Karachi,
Karachi-75270, Pakistan

[3] Faculty of Computing, Engineering and Technology & Centre for Ageing and Mental Health,
Staffordshire University, Stafford, United Kingdom

**Abstract**
Time is one of the most difficult aspects to handle in real world applications such as database systems. Relational database management systems proposed by Codd offer very little built-in query language support for temporal data management. The model itself incorporates neither the concept of time nor any theory of temporal semantics. Many temporal extensions of the relational model have been proposed and some of them are also implemented. This paper offers a brief introduction to temporal database research. We propose a conceptual model for handling time varying attributes in the relational database model with minimal temporal attributes.

## 1. Introduction

The relational model [10] is based on a brand of mathematics called relational algebra. Codd used this concept to manage huge amounts of data very effectively. Codd and others have extended application of this notion to apply to database design. As a consequence they were able to take advantage of the power of mathematical abstraction and the expressiveness of mathematical notation to develop a simple but powerful structure for databases [19].

A relation is a set of tuple(s), and a well-defined structure, called a relational schema. The set of all relation schemas of a database and the database structure is called the database schema. A database schema [19] is a design, which lays the foundation on which the database and the applications are built. A relation has to be in first normal form (FNF) [12][20], which means that the domains of the attributes in its schema may only be of scalar data types. In other words, a relation can be considered as a subset of the Cartesian product of all the attribute domains contained in its schema.

The relational data model only supports functionality to access a single state (the most recent one) of the real world, called a snapshot. The transition from one database state to another (updates) thereby giving up the old state. There exist, however, many application domains which need to have access not only to the most recent state, but also to past and even future states, and the notion of data consistency must be extended to cover all of these database states. Due to the FNF assumption in the relational model, there is a restriction in expressing the data structures. To overcome this drawback, the relational model has been extended to support, non first normal form (NFNF) [12][20] or nested relations [34][43].

Many applications in the real world require management of time varying data such as financial applications, inventory systems, insurance applications, reservation systems, health care management systems, medical databases and decision support systems. Efforts to incorporate the temporal domain into database management system have been ongoing for more than a decade and dozens of temporal models have been proposed [2], [16], [29], [41] and a few of them have been implemented [1], [7], [26],[33], [37], [38].

In the first section we will discuss the important notions of time. It includes the concept of time point or time interval. It is also important to discuss the difference between FNF relations and NFNF relations. Time is used to distinguish between past, present or future states. The recording of time allows the identification when facts are true in the modeled reality (valid time) [12], [20] or when facts are current in the database (transaction time) [12], [20]. Another important concept is time stamping, which can be done with either tuples or attributes in relations.

One area of continuing research interest is the development of a temporal data model capable of representing the temporal dimension of the mini world. The primary focus has been to extend the relational data model to support time-varying information. Second section presents a survey of temporal relational models and proposed extensions to the relational algebra [30], [32] with a classification of the different approaches presented in the literature.

Third section gives an overview of proposed logical temporal data model. Most of the work is related to the semantics and the logical modeling of temporal data [17], [35].

## 2. Temporal Data and Database Types

### 2.1 Relational

A relational database [10], [19] is a set of tables called relations. Each relation contains one or more data categories in columns. Each row contains a unique instance of data for the categories defined by the columns. For example, a typical employee database would include a table that described an employee with columns for name, address, phone number, and so forth. Another table would describe a department: number, name, head, location, and so forth. A user of the database could obtain a view of the database that fits in the user's needs. For example, a human resource manger might need a report about employees who had availed their total leaves on a certain date. A financial services manager in the same company could, from the same tables, obtain a report on accounts that needed to be paid.

### 2.2 Object Relational

An object-relational database (ORD) [19], [27] is a database management system which has the basic features of the relational model with object oriented capabilities such as defining objects, classes and inheritance that are directly supported in database schemas and in the query language. In addition, it supports extension of the data model with custom data-types and methods. ORD permits the modeling of complex data (time varying data) and accommodates various data source formats and standards used by different databases. ORD technology aims to allow developers to raise the level of abstraction at which they view the problem domain.

### 2.3 Object Oriented

An object oriented database [19], [27] is a database model in which information is represented in the form of objects. Time has a complex semantics and object databases have the ability to capture and store it as a basic entity. The typing and inheritance mechanism of object oriented approach gives a better representation of different notion of time.

Object databases also offer some kind of support for representing and managing time varying data using an object query language. Object database and management group (ODMG) made an effort for standardization with the introduction of Object Query Language, OQL.

### 2.4 Temporal Query Languages

Most of the commercial database management systems provide special data types for capturing date, time and timestamps. New version of Oracle also supports user's defined types and user's defined time values. Newer versions of structured query language provide a limited support for querying and managing temporal data. None of the products listed above is said to be a complete temporal database. Essential components for temporal database systems are as follows:
- Temporal data definition language
- Temporal data manipulation language
- Temporal query language [5]
- Temporal constraints(such as temporal referential integrity)

## 3. Discussion

Many data models introduced so far are designed to capture the semantics of temporal data keeping the traditional entity relationship model (ERM) approach. ERM is widely acceptable model for designing conceptual schema of a relational database system. Traditional ERM is capable for capturing the whole temporal aspects. Many extensions [17], [18] have been proposed to extend the ERM in order to capture time varying information in one way or the other. Unified modeling language (UML) [27] is also used as a tool to develop the logical and conceptual schema of the mini world. UML is normally used with the object relational or object oriented databases.

Another important point is how this new conceptual model will be incorporated into a relational database. One approach is to develop a temporal layer and that will be responsible for translating the temporal queries to traditional SQL statements [31]. The other approach is to design a complete temporal query language which will not only support all SQL statements but incorporate new operators based on temporal relational algebra [25]. There are many solutions for this problem and few of them are successfully implemented.

## 4. Important Concepts

Following are some of the important concepts related to the design of temporal model for handling data.

## 4.1 Transaction Time Vs. Valid Time

The transaction time [12], [20], [42] of an object is the time when the object is stored in the database the time at which it is present in the database. For example, if an item is being purchased on 12th June, 2005 recorded on 14th June, 2005 and deleted on 20th Dec, 2005. Another example may be of a company where, an employee receives a pay rise but it comes into effect from the date when payroll clerk enters this salary rise into the database. Transaction time has duration, beginning from insertion and ending at deletion, with multiple insertions and deletions being possible for the same fact.

The valid time [12], [20], [42] of a database object is the time when the object is effective or holds in reality the time when the event took place in reality. For example salary of an employee, whose name is Waleed on 2nd of October, 2001. Valid time is independent of the recording of the fact in a database.

The consideration of valid time is a big problem because some think "only one valid time for every tuple or field" and in some cases there exist many candidate valid times. The actual valid time is always a matter of semantics. The point of view in valid time and transaction time is quite clear and applicable and most of the researchers agree about the significance of these two times in context of the temporal databases [32].

## 4.2 First Normal Form (FNF) Vs. Non-First Normal Form (NFNF)

There are two approaches for handling temporal relations FNF (first normal form) and NFNF (non- first normal form) relations. Relation R is in FNF if and only if all underlying domains contain atomic values only. In contrast, NFNF relations contain multi-values in the time varying attribute. In other words the time varying attribute holds values with sets of intervals. The range of the time stamped attribute is the value domain. In most of the models FNF is used where we have tuple time stamping, whereas attribute time stamping relations can form NFNF relations.

## 4.3 Tuple Time Stamping Vs. Attribute Time Stamping

Temporal models are divided into two broad categories: tuple time stamping and attribute time stamping. In tuple time stamping [20], [42] each tuple is augmented by one or two attributes for the recording of timestamps. One additional attribute can be used to record either the time point at which the tuple becomes valid or the time at which the data is valid. Two additional attributes are used to record the start and stop time points of the corresponding time interval of validity of the corresponding data. Tuple time stamping is usually applied in temporal relational data models supporting only FNF relations.

The alternative is attribute time stamping [20], [42] when the time is associated with every attribute which is time-varying. Note that it is not necessary for every attribute to be time-varying in an attribute time stamping approach. Consequently, a history is formed for each time-varying attribute within each tuple. As a result, the degree of the relation is reduced by one or two compared with the tuple time stamping equivalent relation, since timestamps are part of the attribute values. Attribute time stamping overcomes the disadvantage of data redundancy introduced when applying tuple time stamping. Attribute time stamping adds timestamps to each attribute value. Values in a tuple which are not affected by a modification do not have to be repeated. So, the history of values is stored separately for each attribute.

## 4.4 Time Points Vs. Time Intervals

Extensive study of time representation in temporal databases is conducted by [4], [22], [30], [40]. Three common approaches of time representation are a single time point, an interval and a set of time intervals. In [6], [40], time is expressed using single time points called as events. Most of the temporal models use time intervals to represent time, [23], [28]. A detailed study of point and interval types can be found [45].

There are many reasons that lead to the time point approach. A time point denotes either the start or the end of the lifespan of an object (relation, tuple or attribute). In order to store the whole history of the object, two different attributes need to be added in the relation, i.e. the start point and the stop point, so that the lifespan of that object can be shown. In contrast, time intervals contain the complete information about the lifespan of an object in a "compact style". In the literature, a complete study of intervals has been given, where algebra has been used for their manipulation and the operations defined have proved to be closed. It is also observed that the definition of algebra becomes very difficult when time intervals [26] are used instead of time points [44][45].

## 5. Temporal Data Models

Since most of the work in the research area of temporal databases has been done with respect to the relational data model, numerous proposals can be found some of which are already implemented. Table 1 mentions some of the most important ones with respect to the work presented in this paper. Temporal data models can be categorized as tuple stamped or attribute time stamped, FNF or NFNF, valid time or transaction time and some other important attributes.

Table 1. Summary of temporal extensions of the relational data model

| Temporal Model | NF | Time Stamping | Time | Features / Query Language |
|---|---|---|---|---|
| Ariav's model [2]. | 1NF | Tuple | VT & TT | Time stamps are based on time points |
| Ben-Zvi's Model [3] | | Tuple | VT & TT | Time Relational model Extension of snapshot algebra |
| Clifford & Croker's Model [6], [7] | NINF | Attribute | VT | HDBM Inhomogeneous |
| Gadia's model [13][14] | NINF | Attribute | VT | Homogeneous HRQUEL |
| Gadia & Yeung's Model [16] | | Attribute | VT & TT | Heterogeneous model |
| Jensen & Snodgrass's Model [21][22] | 1NF | Tuple | VT & TT | Bi-temporal Conceptual Model (BCDM) |
| Lorentzos's Model [24] | 1NF | Attribute | VT | Interval Relational Model (IRM) Extended Relational Model (XRM) |
| McKenzie's Model[28],[29] | NINF | Attribute | VT & TT | Extension of Snapshot algebra Historical algebra |
| Snodgrass's Model [37] | NINF | Tuple | VT & TT | TQUEL |
| Tansel's Model [20] | NINF | Attribute | VT | Non homogeneous |
| TSQL2 [38] | 1NF | Tuple | VT, TT & User Defined Time | Homogeneous model |

## 6. Methodology

We adopt comparison methodology to propose new model of temporal database. This proposed model is based on comparison of same level but different behavior of different available databases such as RDBMS and OODBMS.

## 7. Introduction to the Proposed Model

By now the relational model is the most effective method for organizing huge amounts of data and still the widely accepted technology amongst vendors and enterprises all around the world. Temporal data has its own semantics and organization [15] of such data requires some modifications in the relational model [4]. A lot of work has been done in the area of temporal models (fig. 1). Attributes in a relation can be time varying or non-time varying attributes. For e.g. salary attribute in an employee relation is a time varying relation, because salary changes with time and contrary gender attribute in the employee relation is a non time varying attribute.

7.1 Structure of Time:

The structure of the set time to be $< T,< >$, where T is some countable set and less than, "<" is a linear order on T, i.e. for any two time points t1 and t2, either t1 = t2, t1 $\leq$ t2, or t2 $\leq$ t1. In our model we treat time as discrete and isomorphic to the natural numbers because any practical domain that we might define for time attributes in our proposed model would have at most an infinite countable set of names for time moments or time intervals.

Tuple time-stamping approach has been adopted to define a temporal model. The reason for this is the simplicity and to keep the 1NF assumption and the essence of the relational model.

The proposed logical temporal data model (fig. 1) comprises of three main constructs namely, the entity (temporal or no temporal), attributes (time-varying, non-time-varying and partial time varying attributes) and thirdly the relationship type amongst the entities are n-ary and categorized as temporal, non-temporal or semi temporal. Fully temporal attribute will be handled by a separate entity called temporal entity.

Following are some constraints to ensure the consistency of the conceptual schema:

- Entity is categorized as a temporal and non-temporal entity
- Temporal entity type must have a combinational primary key composed of time-varying and non-time-varying attributes. The activation start time is the part of the key.
- The n-ary relationship determines whether it is non-temporal, temporal or semi temporal relationship.

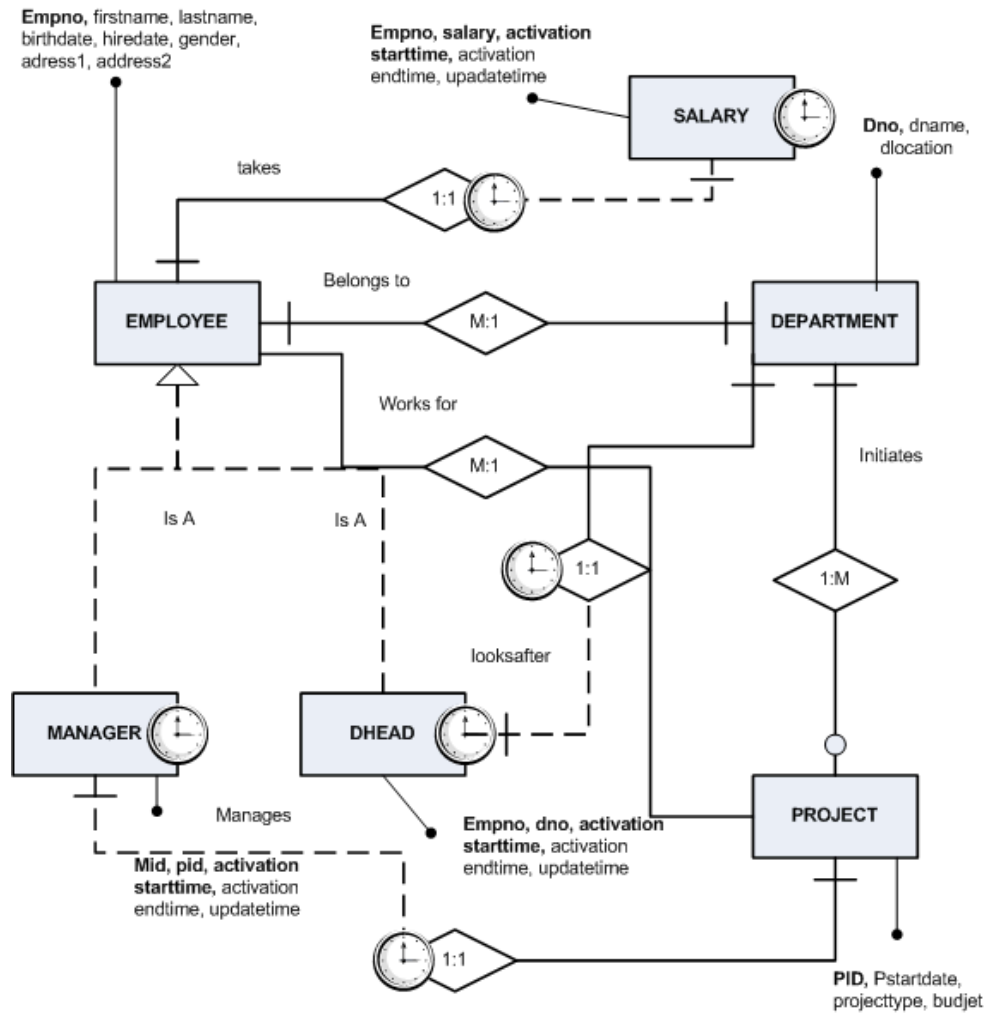

Figure 1: Temporal Conceptual Model of Employee Database

### 7.2 Representing Temporal Characteristics

- We have introduced a new term as activation_start and activation_end rather than valid time because validity itself is of many types and it creates confusion while defining an entity.
- It represents temporal facts both at time points as well as on intervals.
- Update time is introduced instead of transaction time. An update refers to change in data (tuple) of any sort (insert, delete or change).
- Activation time can be represented with different time granularities [11] such as year, month, week, day, hour, minute and second and even beyond that. Conversion from one time granularity [46] to another is accomplished by the conversion functions.

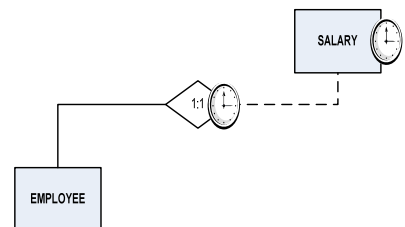

Figure 2(a): Employee and Salary Relationship

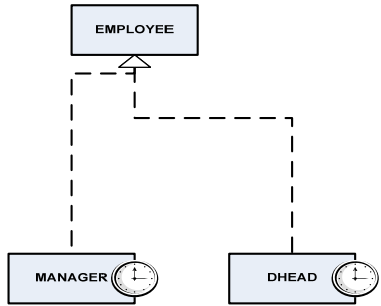

Figure 2(b): Generalized-Specialized Employee Relationship with Manager and Dhead

Salary is a temporal entity represented in fig. 2(a) that records the history of an employee salary. Activation_start shows that the value associated to it is the starting point. Employee represents the snapshot of employee's entity and it is non-temporal, but it has a relationship with the entity salary which is temporal in nature. A dotted line from a salary is representing partial temporal relationship (see fig. 2(a). Manager and dhead are special type of employee represented in fig. 2(b) where manager and dhead are temporal entities.

### 7.3 Temporal Relation Schema

A temporal relation schema TR = < A, K > is an ordered pair consisting of finite set of A = {A1, A2,…,An, activation_start, activation_end, updatetime}. Each temporal relation has a current state, and the changed states (future and past). Relations and key attributes of the EMPLOYEE database are as follows.

Employee = (Empno, firstname, lastname, birthdate, hiredate, address1, address2, gender)
- Empno is a primary key and it is not a time varying attribute
- Address is a partial time varying attribute: handled by an application program

Manager = (empno, activation_start, activation_end, updatetime)
- Manager is a temporal relation
- Empno, startdate is a time-variant key
- Pid is a foreign key referencing project schema

Dhead = (empno, dno, activation_start, activation_end, updatetime)
- Dhead is a temporal relation
- Empno and startdate is a combinational primary key
- Dno is foreign key referencing department schema

Department = (Dno, dname, dlocation, dcontact)
- Dno is a primary key and it is not a time varying attribute
- No temporal attributes

Project = (Pid, pstartdate, ptype, budget)
- Pid is a primary key and it is not a time varying attribute
- No temporal attributes
- Dno is a foreign key referencing department schema

Salary: (empno, salary, activation_start, activation_end, updatetime)
- Salary is a temporal schema
- Primary key is (Empno, salary and activation_start)
- FK (Employee_Empno)

Project: (Pid, pstartdate, projecttype, budget)
- Project is a non-temporal schema
- Primary key is Pid

Table 2a: EMPLOYEE

| Empno | First name | Last name | Hiredate | Birthdate | Address1 | Aderess2 | Gender |
|---|---|---|---|---|---|---|---|
| 101 | ahmed | mumtaz | 01-03-2000 | 02-05-1980 | A60 | Street 27 | M |
| 102 | majid | bhati | 14-03-2002 | 10-02-1976 | A78 | Street 22 | M |
| 103 | aliya | rasheed | 13-05-1995 | 11-08-1968 | C12 | Street 21 | F |
| 104 | salim | khan | 10-04-2006 | 12-11-1981 | B21 | Street 11 | M |
| 105 | danish | khan | 11-07-1997 | 01-02-1978 | C76 | Street 9 | M |
| 106 | sajid | mahmood | 02-04-1999 | 10-09-1977 | D1 | Street 9 | M |
| …. | …. | …. | …. | …. | … | … | …. |

Table 2b: DHEAD (Temporal relation)

| Employee _Empno | Department_ Deptno | Activation _start | Activation _end | Update time |
|---|---|---|---|---|
| 103 | 10 | 13-05-1995 | 13-05-1998 | 14-05-1995 |
| 103 | 10 | 14-05-1998 | 21-07-2002 | 14-05-1998 |
| 103 | 20 | 22-07-2002 | Current time | 22-07-2002 |
| …. | …. | …. | …. | …. |

Table 2c: SALARY (Temporal relation)

| Empno | Salary | Activation _start | Activation _end | Update time |
|---|---|---|---|---|
| 103 | Rs. 900 | 13-05-1995 | 01-03-1998 | 12-05-1995 |
| 103 | Rs. 1100 | 02-03-1998 | 11-12-2002 | 27-02-1998 |
| 103 | Rs. 1300 | 12-12-2002 | 10-07-2005 | 11-12-2002 |
| 103 | Rs. 1500 | 11-07-2005 | Current time | 09-07-2005 |
| …. | …. | …. | …. | …. |
| …. | …. | …. | …. | …. |

The non temporal facts of employees are stored in a relation employee (table 2a). Although the address attribute is also a time varying attribute, employee do not change address so often so it can be managed in the same relation, introducing a separate relation over here will increase overhead. Dhead relation (fig. 2b) stores the current and historic information of the head of the department.

Current time is a function which returns the current time. For e.g. employee 103 is getting Rs. 1500 since 11th of July, 2005. Minute is the default time granularity [39][46] for defining these relations but it can be changed by invoking conversion functions if required. The complete salary history of the employee 104 can easily be recovered from the salary relation (table 2c).

Update operations (insert, delete and change) result in insertion of a new tuple that contains the new values associated with a new timestamp. The new attribute value does not replace the previous (old) value of the attribute in case of an update such as deletion or insertion. Similarly, the deletion of tuples does not cause an actual deletion from database tables.

## 8. Conclusions

Temporal data model provides the mechanism to capture the time varying nature of entities and to design temporal query language to retrieve, manipulate and process the temporal data. The semantics of time is a big issue to deal with and it varies with the application and domain. Temporal relational database also holds time series data about entities and the events occurring in real time.

The proposed conceptual temporal relational model can be easily extended to incorporate time varying attributes. In our model each time varying attribute, identified in the mini world has a separate relation called as temporal relation. The concept of temporal relationship is introduced. We have presented some of the most important temporal models and discussed the semantics of our proposed temporal schema of the employee database.

The benefits of the proposed model are not restricted to reducing temporal attributes in the relation but it keeps the essence of the relational model and causes less overhead [36]. Due to the simplicity of the conceptual model, it will be easier to implement. Inclusion of too many temporal attributes in relations causes inconsistencies and redundancy in the database [36].

## 9. Future work

There are still some points which have not been addressed in this paper. The overhead reduced by this model in the case of insert, update and change operations should formally be estimated and verified through experimental results. The fact that some events might occur in the modeled reality, but the reflection of these events to the database is delayed. We defer the discussion of these points in a future work.

Temporal aspect is significantly important in various fields and it can be very helpful in analyzing and understanding domain performance. We will evaluate this model in healthcare mobile system for the analysis of patients' movements to improve healthcare service. One of our co-author is working on healthcare care application domain in which context based knowledge management system model will be designed to improve service for hospital patients. In this model knowledge factor in changing patients' movement scenario will be captured through mobile technology. Our temporal model will be of vital use for their database implementation and integration.

IJCSI International Journal of Computer Science Issues, Vol. 7, Issue 1, No. 1, January 2010
www.IJCSI.org

9[46] X. S. Wang, G. Mason, Logical Design for Temporal Databases with Multiple Granularities University, Fairfax, ACM 0362-5915/97/0600–0115 ACM Transactions on Database Systems, Vol. 22, No. 2, (1997).**Mahmood, Nadeem**
Nadeem Mahmood is a PhD research scholar and an assistant professor in Department of Computer Science, (UBIT), University of Karachi. His area of research is the application of temporal logic in relational database systems, temporal databases and temporal data modeling. He has masters in Computer Science from University of Karachi. He has several research publications in journals and presented papers in national and international conferences. Currently a member of ACM and member of Healthcare Care Information System group at University of Karachi.

**Burney, Aqil**
Dr. S. M. Ail Burney received his B.Sc, first class first M.Sc. M.Phil. from Karachi University in 1970, 1972 and 1983 respectively. He received Ph.D. degree in Mathematics from Strathclyde University, Glasgow with specialization in estimation, modeling and simulation of multivariate Time series models using algorithmic approach with software development. He is currently professor at (UBIT) University of Karachi and approved supervisor in Computer Science and Statistics by the High Education Commission, Government of Pakistan. His research interest includes AI, soft computing, neural networks, fuzzy logic, data mining, statistics, simulation and stochastic modeling of mobile communication system and networks and network security. He is author of three books, various technical reports and supervised more than 100 software/Information technology projects of Masters level degree programs and project director of various projects funded by Government of Pakistan. He is member of IEEE (USA), ACM (USA) and fellow of Royal Statistical Society United Kingdom and also a member of Islamic society of Statistical Sciences. Dr. Burney has received appreciations and awards for his research and as educationist including NCR-2002 award for Best National Educationist.

**Ahsan, Kamran**
Kamran Ahsan is a PhD researcher and lecturer in FCET (Faculty of Computing, Engineering and Technology) and, web researcher in Centre for Ageing and Mental Health, Staffordshire University, UK since 2005.  He is researcher with Health Strategic Authority, UK. He has an MSc in Mobile Computer Systems from Staffordshire University and MCS in Computer Science from University of Karachi. He has several publications and member of various Journals/Conferences reviewing committee. His research interests are mobile technology applications in healthcare including knowledge management.www.IJCSI.org